\documentclass[fleqn,usenatbib]{mnras}

\usepackage[T1]{fontenc}

\DeclareRobustCommand{\VAN}[3]{#2}
\let\VANthebibliography\thebibliography
\def\thebibliography{\DeclareRobustCommand{\VAN}[3]{##3}\VANthebibliography}

\usepackage{graphicx}	
\usepackage{amsmath}	
\usepackage{subcaption}
\usepackage{caption}

\usepackage{amsmath,amssymb,gensymb,times,graphicx,morefloats,xcolor}
\usepackage{amsmath}
\usepackage{graphics,graphicx,url,verbatim,xspace}

\usepackage{changepage}
\usepackage{mathtools} 
\usepackage{comment}
\usepackage{gensymb}
\usepackage{hyperref}
\usepackage[export]{adjustbox}

\usepackage{newtxtext,newtxmath}

\usepackage{ulem}

\newcommand{\changed}[1]{\textcolor{black}{#1}}

\newcommand{\Msun}{\mbox{M$_\odot$}}
\newcommand{\Mjup}{\mbox{M$_{\rm Jup}$}}

\newcommand{\HIPA}{HIP~67506~A}
\newcommand{\HIPC}{HIP~67506~C}
\newcommand{\HIPB}{HIP~67506~B}
\newcommand{\lod}{$\lambda$/$D$}

\usepackage{amsmath,amssymb,gensymb,times,graphicx,morefloats,color}
\usepackage{amsmath}
\usepackage{graphics,graphicx,url,verbatim,xspace}

\usepackage{changepage}
\usepackage{mathtools} 
\usepackage{comment}
\usepackage{gensymb}
\usepackage{hyperref}

\usepackage{enumitem}

\graphicspath{{./}{figures/}}

\title[HIP 67056 AC]{HIP 67506 C: MagAO-X Confirmation of a New Low-Mass Stellar Companion to HIP 67506 A }

\author[L. A. Pearce et al.]{
Logan A. Pearce$^{1,2}$\thanks{E-mail: loganpearce1@arizona.edu},
Jared R. Males$^{1}$,
Sebastiaan Y. Haffert$^{1,10}$,
Laird M. Close$^{1}$,
\newauthor{Joseph D. Long$^{1}$,
Avalon L. McLeod$^{1,3}$,
Justin M. Knight$^{1,3}$,
Alexander D. Hedglen$^{1,3}$,}
\newauthor{
Alycia J. Weinberger$^{4}$,
Olivier Guyon$^{1,3,5,6}$,
Maggie Kautz$^{1,3}$,
Kyle Van Gorkom$^{1,3}$,}
\newauthor{Jennifer Lumbres$^{1,3}$,
Lauren Schatz$^{8}$,
Alex Rodack$^{1,3}$,
Victor Gasho$^{1}$,}
\newauthor{Jay Kueny$^{1,3}$,
Warren Foster$^{1,3}$,
Katie M. Morzinski$^{1}$,
Philip M. Hinz$^{9}$
}
\\
$^{1}$Steward Observatory, University of Arizona, Tucson, AZ 85721, USA\\
$^{2}$NSF Graduate Research Fellow\\
$^{3}$James C. Wyant College of Optical Sciences, University of Arizona, 1630 E University Blvd, Tucson, AZ
85719, USA\\
$^{4}$Earth and Planets Laboratory, Carnegie Institution for Science, 5241 Broad Branch Road NW, Washington, DC 20015-1305\\
$^{5}$National Astronomical Observatory of Japan, Subaru Telescope, National Institutes of
Natural Sciences, Hilo, HI 96720, USA\\
$^{6}$Astrobiology Center, National Institutes of Natural Sciences, 2-21-1 Osawa, Mitaka, Tokyo,
JAPAN\\
$^{8}$Kirtland Air Force Base, Air Force Research Laboratory, Albuquerque, NM, USA\\
$^{9}$UC Santa Cruz, 1156 High St, Santa Cruz CA 95064, USA\\
$^{10}$NASA Hubble Fellow
}

\date{Accepted XXX. Received YYY; in original form ZZZ}

\pubyear{2022}

\begin{document}
\defcitealias{Baraffe2015BHAC}{BHAC15}
\label{firstpage}
\pagerange{\pageref{firstpage}--\pageref{lastpage}}
\maketitle

\begin{abstract}

We report the confirmation of HIP~67506~C, a new stellar companion to HIP~67506~A.  We previously reported a candidate signal at 2$\lambda$/D (240~mas) in L$^{\prime}$ in MagAO/Clio imaging using the binary differential imaging technique.  Several additional indirect signals showed that the candidate signal merited follow-up: significant astrometric acceleration in Gaia DR3, Hipparcos-Gaia proper motion anomaly, and overluminosity compared to single main sequence stars. We confirmed the companion, HIP~67506~C, at 0.1" with MagAO-X in April, 2022.  We characterized HIP~67506~C MagAO-X photometry and astrometry, and estimated spectral type K7-M2; we also re-evaluated HIP~67506~A in light of the close companion.  Additionally we show that a previously identified 9" companion, HIP~67506~B, is a much further distant unassociated background star.  We also discuss the utility of indirect signposts in identifying small inner working angle candidate companions.
\end{abstract}

\begin{keywords}
planets and satellites: detection, (stars:) binaries: visual, stars: statistics, methods: data analysis, methods: observational
\end{keywords}



\section{Introduction}\label{sec:introduction}


High-contrast imaging searches have found very low occurrence rates for close substellar companions. \changed{For example, $9^{+5}_{-4}$\% for 5-13~\Mjup, $\sim0.8^{+0.8}_{-0.5}$\% for 13-80~\Mjup\ companions within 10-100 AU in the recent results from the Gemini Planet Imager Exoplanet Survey (GPIES); \citep{nielsen_gemini_2019}, while the SHINE survey \citep{Vigan2021SPHEREOccurrenceRates} found frequency of systems with at least one substellar companion to be 23.0$^{+13.5}_{-9.7}$\%, 5.8$^{+4.7}_{-2.8}$\%, and 12.6$^{+12.9}_{-7.1}$\% for BA, FGK, and M stars}. Yet radial velocity, transit, and microlensing surveys point to higher occurrence rates in regions promising for future direct imaging contrasts and separation  \citep[e.g.][]{Bryan2019JupiterAnalogs,Herman2019LongPeriodTransit,Poleski2021MicrolensingWidePlanetsCommon}. Decreasing the effective inner working angle (IWA) of observations increases the area of the accessible region proportional to (IWA)$^{-2}$. Smaller IWAs extend the reach to tighter regimes of nearby stars, and to the planetary regime of more distant stars \citep{Mawet2012ReviewCoronographicMethods}. Working at small IWAs will be vital for the future of the high-contrast imaging field.

\changed{\citealt{rodigas_direct_2015} demonstrated that for visual binaries of separation $\approx$2 -- 10\arcsec\ and approximately equal magnitude, a starlight subtraction via a principal component analysis-based reference differential imaging (RDI) algorithm using each star of the binary as reference for the other -- termed binary differential imaging (BDI) -- outperforms the common angular differential imaging technique at close separations. In \citealt{Pearce2022BDI} we used BDI to reduce a set of 17 visual binaries imaged in L$^\prime$ and 3.95$\mu$m filters with MagAO/Clio instrument on the Magellan Clay Telescope at Las Campanas Observatory from 2015-2017. In that work we reported detection of a candidate companion signal at 2\lod\ separation to the star \HIPA. Due to the proximity to the star's core we were unable to determine the nature of the companion, but had evidence to suggest it might be near the stellar/substellar mass boundary.
}

\changed{In this work we report the results of follow-up observations of \HIPA\ with the MagAO-X instrument on the Magellan Clay telescope in April 2022 to confirm the candidate signal.}  We report the discovery of \HIPC, a previously unknown early-M type 0.1\arcsec\ ($\sim20$ AU) companion to \HIPA.  
\changed{In Section \ref{sec:stellarproperties} we describe the indirect indications pointing to the existence of a hidden companion.} In Section \ref{sec:observationsandanalysis} we describe our MagAO-X follow up observations and confirmation of \HIPC, and in Section \ref{sec:results} our astrometric and photometric characterization. \changed{Additionally in Appendix A we demonstrate that the previously identified 9\arcsec-separated star \HIPB\ is not actually physically associated. }

\section{Stellar Properties}\label{sec:stellarproperties}
\HIPA\ is a field star (99.9\% probability in BANYAN $\Sigma$; \citealt{gagne_banyan_2018}) at 221.6$\pm$1.8 pc \citep{gaiaEDR3}.  It was identified as type G5 \citep{SpencerJones1939SpT}, mass 1.2\Msun\ \citep{Chandler2016HabitableZones}, with effective temperature T$_{\rm{eff}}$~=~6077~$\pm$~150~K and luminosity L~=~0.37~$\pm$~0.07~L$_\odot$ \citep{mcdonald_fundamental_2012}.  In \citet{Pearce2022BDI} we used these values to estimate an age of $\approx$200~Myr from isochrone fitting to \cite{Baraffe2015BHAC} isochrones. It was identified in the Hipparcos and Tycho Doubles and Multiples Catalog \citep{Hip1997Vizier} as a binary system with another star (\HIPB) with separation 9\arcsec, and dubbed HIP~67506~A and B.

\subsection{Indicators of a companion to \HIPA}\label{sec:bdi}

In \citet{Pearce2022BDI} we observed 17 visual binary systems and reduced the images using the Binary Differential Imaging (BDI) technique \citep[see also ][]{rodigas_direct_2015} with Magellan Adaptive Optics system (MagAO) \citep{close_diffraction-limited_2013} and Clio science camera on the Magellan Clay Telescope at Las Campanas Observatory in MKO L$^\prime$ and 3.95$\mu$m filters, from 2014--2017.  To summarize briefly, we simultaneously observed a science and PSF reference target by selecting binaries of nearly equal magnitude, separated enough that their PSF features do not overlap, but close enough to be within the isoplanatic patch at these wavelengths, making the target and reference PSF as close to equal in structure and signal-to-noise ratio as possible.  We then reduced each star with the other as the PSF reference, using Karhunen-Lo\`eve Image Projection \citep[KLIP; ][]{soummer_detection_2012} to reconstruct a model PSF from the reference star to subtract from the target star.

\begin{figure}
\centering
\includegraphics[width=0.48\textwidth]{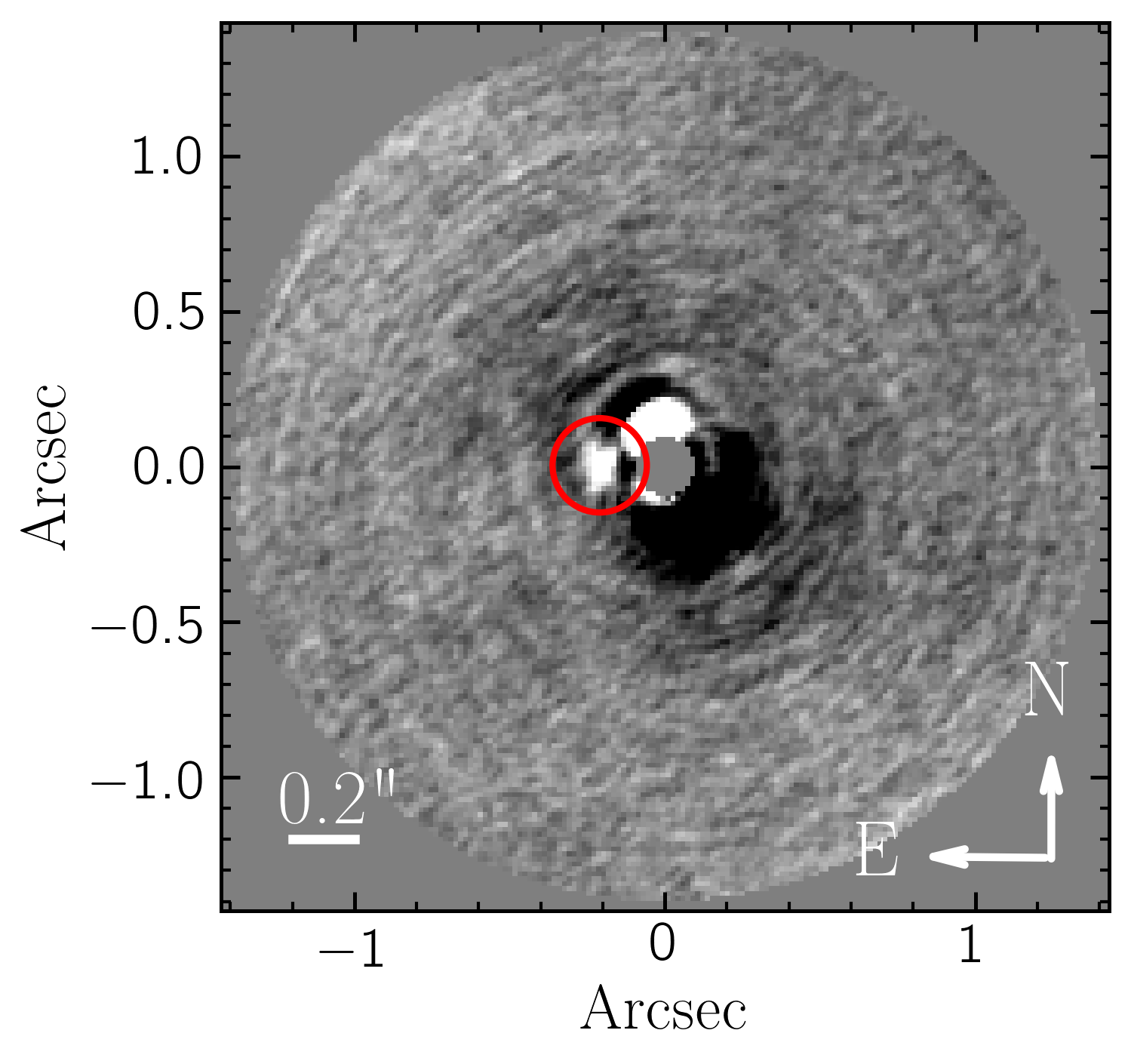}\\
\caption{MKO L$^\prime$ KLIP-reduced image of \HIPA\ from our Binary Differential Imaging survey described in \citet{Pearce2022BDI}.  The central star is masked in the reduction, and the candidate signal is marked with a red circle $\sim$2\arcsec\ (2.0~\lod) to the east.  This was identified as a candidate signal due to the fact that it did not appear to smear azimuthally with derotation like the other residual structures at similar separation, and the other indications described in Section \ref{sec:bdi}
}
\label{fig:bdi-hipa}
\end{figure}

We observed \HIPA B on 2015-05-31 as part of this survey and detected a candidate companion signal $\sim$0.2\arcsec\ East of \HIPA.  Figure \ref{fig:bdi-hipa} displays the KLIP-reduced image of \HIPA\ from that paper, with the candidate signal marked by the red circle.  The candidate signal is distorted from a typical PSF shape -- due its proximity to the star's core (at $2\lambda$/D) the signal was corrupted by PSF subtraction.  However the fact that it did not appear to smear azimuthally like the other residuals at that same separation points to the possibility of its being a true companion signal.  

\begin{figure}
\centering
\includegraphics[width=0.48\textwidth]{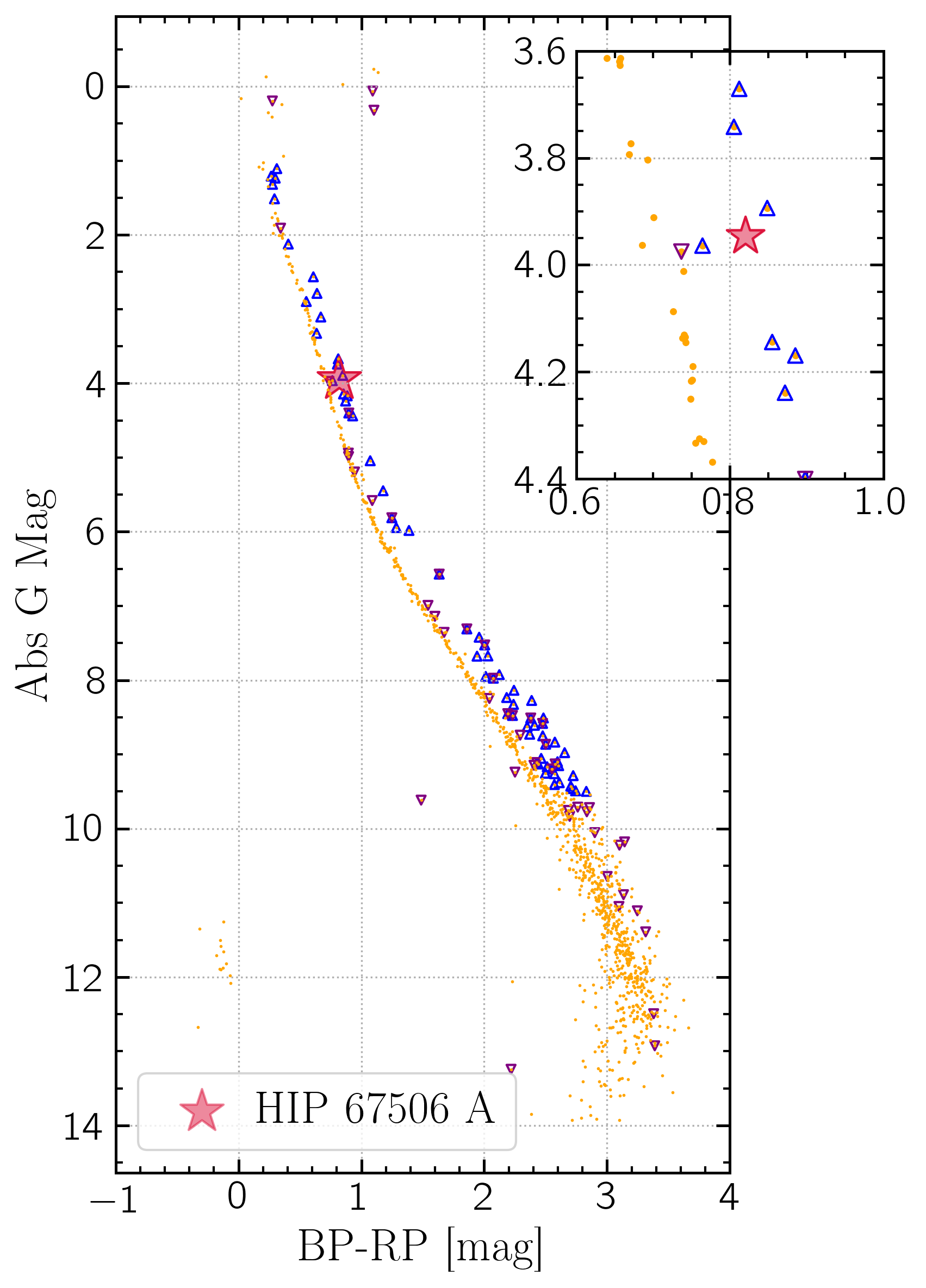}\\
\caption{Gaia EDR3 BP minus RP vs absolute G magnitude color-magnitude diagram of Praesepe Cluster members identified in \citealt{deacon_wide_2020} (orange).  Objects they flagged as possible overluminous binaries are outlined in blue up-pointing triangles, and purple down-pointing triangles are objects they flagged with elevated astrometric noise, following their Figure 4.  The position of HIP 67506 is marked with a red star in the main and inset axis, which shows a close view of the region surrounding \HIPA.  \HIPA\ falls on the overluminous region above the main sequence, pointing to the presence of an unresolved stellar companion.
}
\label{fig:praesepe_cmd}
\end{figure}

\begin{figure*}
\centering
\includegraphics[width=0.9\textwidth]{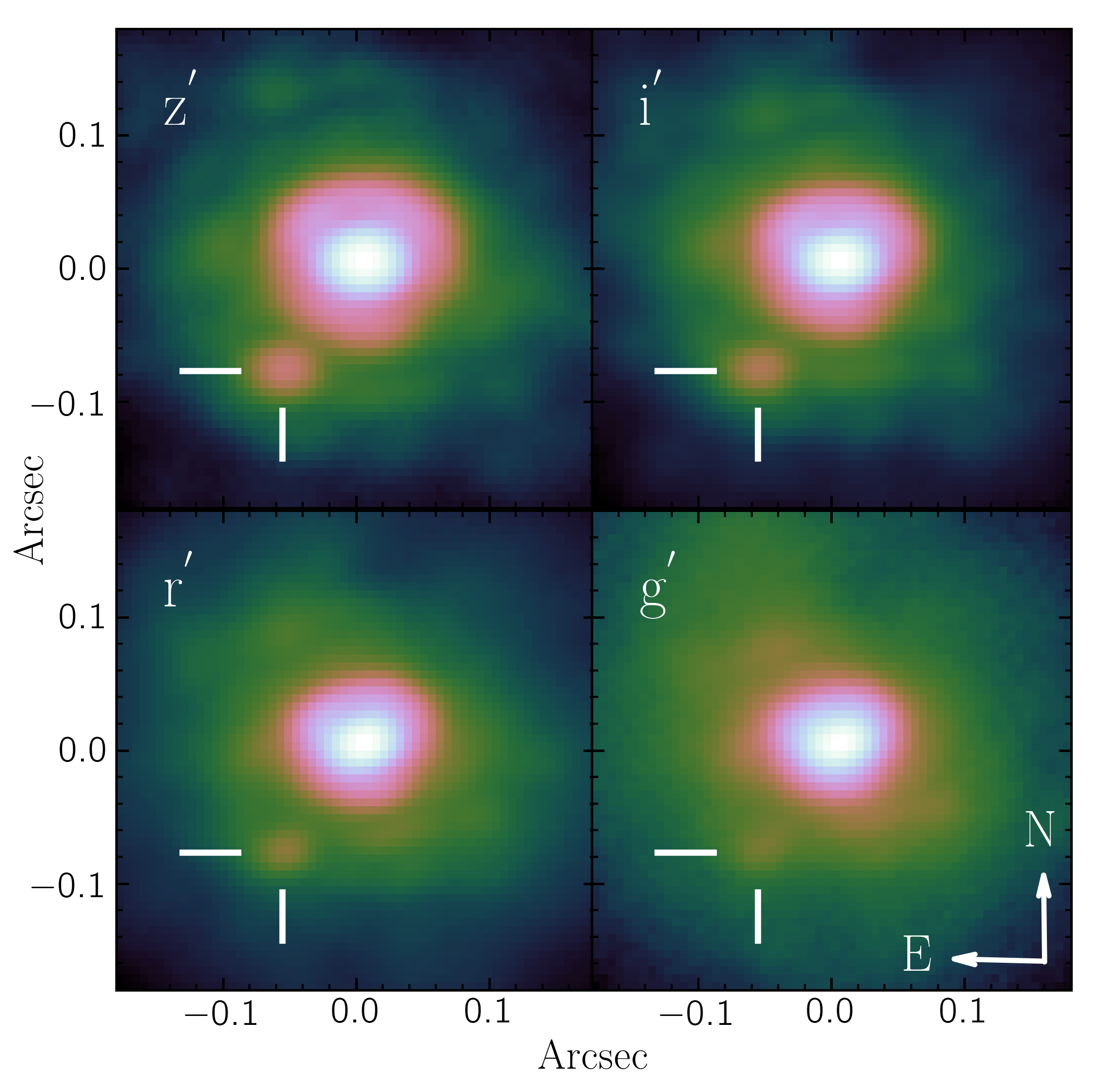}\\
\caption{MagAO-X images of \HIPA and \HIPC in the four photometric filters g$^\prime$, r$^\prime$, i$^\prime$, z$^\prime$, shown with log stretch.  \HIPA is centered in each image, and \HIPC, located 0.1\arcsec\ to the south east, is marked by the white pointers.  North is up and East is left, and the stretch and spatial scale is same for each image.
}
\label{fig:four-filter-image}
\end{figure*}

There are secondary indications of a companion to \HIPA.  Figure \ref{fig:praesepe_cmd} shows a Gaia EDR3 BP minus RP vs absolute G magnitude color-magnitude diagram of Praesepe Cluster members identified in \citealt{deacon_wide_2020} (orange), reproducing their Figure 4.  Members they flagged as overluminous and with elevated astrometric noise in Gaia ERD3, indicating an unresolved companion, are marked with blue and purple triangles respectively. \HIPA\ is marked with a red star in the main and inset axes.  \HIPA\ clearly falls on the overluminous region above the main sequence, indicating that the flux measured by Gaia is abnormally high for a single star, pointing to the presence of an unresolved stellar companion. 

\HIPA\ also has indicators of an unresolved companion in Gaia astrometry.  The Gaia Renormalized Unit Weight Error (RUWE) is a signpost for unresolved companions. RUWE encapsulates all sources of error in the fit to the assumed single star astrometric model, corrected for correlation with source color and magnitude. RUWE~$\approx$~1 is expected for a well-behaved solution \citep{lindegren_re-normalising_2018}\footnote{\url{https://www.cosmos.esa.int/web/gaia/dr2-known-issues\#AstrometryConsiderations}}. RUWE $>$2 indicates signficant devation from a single star model. \HIPA\ has RUWE$= 2.02$ in Gaia EDR3, indicating that a companion is likely.  

While RUWE is the most complete and easy to interpret metric \citep{lindegren_re-normalising_2018}, other metrics in Gaia can probe multiplicity. Perturbations of the source photocenter (caused by orbiting unresolved objects) compared to the center-of-mass motion (which moves as a single star) will cause the observations to be a poor match to the fitting model, which registers as excess noise via the \texttt{astrometric\_excess\_noise} parameter, and whose significance is captured in the \texttt{astrometric\_excess\_noise\_sig} parameter ($>$2 indicates significant excess noise). The \texttt{astrometric\_chi2\_al} term reports the $\chi^2$ value of the observations to the fitting model, with lower values indicating better fit to observations. From the image parameter determination (IPD) phase, \texttt{ipd\_gof\_harmonic\_amplitude} is sensitive to elongated PSF shapes relative to the scan direction (larger values indicate more elongation), and \texttt{ipd\_frac\_multi\_peak} reports the percentage of observations which contained more than one peak in the windows\footnote{See \url{https://gea.esac.esa.int/archive/documentation/GEDR3/Gaia_archive/chap\_datamodel/sec\_dm\_main\_tables/ssec\_dm\_gaia\_source.html} for complete description of Gaia catalog contents}.

\begin{table}
	\centering
	\caption{Multiplicity Metrics for \HIPA}
	\label{tab:gaia-metrics}
	\begin{tabular}{lc} 
		\hline
		Metric & Value\\
		\hline
		\multicolumn{2}{c}{Gaia}\\
		\hline
		RUWE & 2.02 \\
		\texttt{astrometric\_excess\_noise} & 0.22 \\
		\texttt{astrometric\_excess\_noise\_sig} & 75.16 \\
		\texttt{astrometric\_chi2\_al} & 2277.97 \\
		\texttt{ipd\_gof\_harmonic\_amplitude} & 0.0099 \\
		\texttt{ipd\_frac\_multi\_peak} & 0 \\
		\hline
		\multicolumn{2}{c}{Hipparcos-Gaia Accelerations}\\
		\hline
		HGCA $\chi^2$ \citep{Brandt2021_HGCA_ERD3} & 41 \\
		M$_2$ at 23AU from from PMa \citep{Kervella2022} & 270~\Mjup\\
	\end{tabular}
\end{table}

Table \ref{tab:gaia-metrics} shows values of these metrics for HIP 67506 A. The IPD parameters are small and insignificant, suggesting that there are no marginally resolved sources ($\rho\sim$0.1-1.2\arcsec, separation larger than the resolution limit but smaller than the confusion limit,   \citealt{gaiaEDR3}) present in the images, however the astrometric noise parameters are large and significant, affirming the presence of subsystems. This points to a companion near or below the resolution limit of $\approx$0.1\arcsec.

Finally, \HIPA\ also shows significant acceleration between the Hipparcos and Gaia astrometric measurements.  The Hipparcos-Gaia Catalog of Accelerations \citep[HGCA; ][]{Brandt2021_HGCA_ERD3} measures the change in proper motion between a star's Hipparcos and Gaia proper motion measurements, as well as the positional difference between the missions, divided by the $\sim$24 year time baseline, and quantifies the deviation from linear motion.  This acceleration is called the proper motion anomaly (PMa).  The HGCA shows a significant PMa for \HIPA, with a $\chi^2 = 41$ for the goodness of fit of a linear proper motion to the measured astrometry.  This points to unresolved subsystems causing acceleration.  

Additionally, \citealt{Kervella2022} produced a PMa catalog for Hipparcos-Gaia EDR3 which also shows significant acceleration for \HIPA\ (S/N = 9.31). They used the measured tangential velocity anomaly to constrain the mass of the object causing acceleration (which is degenerate with separation; \citealt{kervella_stellar_2019}).  Using a mass of 1.3~\Msun\ for \HIPA, they estimate a companion of mass 180~\Mjup\ at 10~au causing the observed acceleration of \HIPA.  Extrapolating this out to the 2015 projected separation of \HIPC\ (48~AU), the acceleration would be caused by a $\sim$400~\Mjup\ object. \changed{The position angle of the acceleration given in \citealt{Kervella2022} is 96.6$\pm$3.8$^\circ$ for the 2016.0 Gaia epoch, which agrees within uncertainty with the candidate signal position angle in 2015.4, as would be expected if the candidate signal were the cause of the observed acceleration.
}

Combined with the candidate signal in our 2015 MagAO observation, these other lines of evidence point to a strong chance of this being a genuine companion signal which merited follow-up for confirmation and characterization.  

\begin{table}
	\centering
	\caption{Stellar Properties of \HIPA}
	\label{tab:HIPA-properties-old-new}
	\begin{tabular}{llcl} 
		\hline
		Parameter & Previous Value & Ref & Our Value\\
		\hline
		Distance \changed{(pc)} & 102$\pm$86 & 1 & 221.6$\pm$1.8$^{a}$\\
		Mass \changed{(\Msun)} & 1.2$\pm$0.1 & 2 & 1.2$\pm$0.2\\
		Spectral Type & G5 & 3 & F8--G2\\
		T$_{\rm{eff}}$ \changed{(K)} & 6077~$\pm$~150 & 4 & 6000$\pm$350\\
		Luminosity \changed{(L$_\odot$)} & 0.37~$\pm$~0.07 & 4 & 1.91$^{+0.28}_{-0.32}$\\
		Sloan m$_{g\prime}$ & 11.04$\pm$0.01 & 5 & 11.04$\pm$0.01\\
		Sloan m$_{r\prime}$ & 10.66$\pm$0.01 & 5 & 10.67$\pm$0.01\\
		Sloan m$_{i\prime}$ & 10.56$\pm$0.01 & 5 & 10.59$\pm$0.01\\
		Sloan m$_{z\prime}$ & 10.50$\pm$0.01 & 5 & 10.55$\pm$0.01\\
		Sloan g-r & 0.38$\pm$0.02 & 5 & 0.37$\pm$0.02\\
		Sloan r-i & 0.11$\pm$0.02 & 5 & 0.09$\pm$0.02\\
		\hline
		\multicolumn{4}{l}{(1) \citealt{vanLeeuwen2007HipReduction}, (2) \citealt{Chandler2016HabitableZones},}\\
		\multicolumn{4}{l}{(3) \citealt{SpencerJones1939SpT}, (4) \citealt{mcdonald_fundamental_2012},}\\
		\multicolumn{4}{l}{(5) \cite{UCAC4-Vizier}, $^{a}$Gaia EDR3 \cite{gaiaEDR3}}
	\end{tabular}
\end{table}

\begin{table}
	\centering
	\caption{Properties of \HIPC}
	\label{tab:HIPC-properties}
	\begin{tabular}{ll} 
		\hline
		Parameter & Value\\
		\hline
		\multicolumn{2}{c}{Stellar Properties}\\
		\hline
		\vspace{0.1cm}
		Spectral Type & K7--M2 \\
		\vspace{0.13cm}
		T$_{\rm{eff}}$ & 3600$^{+250}_{-350}$~K \\
		\vspace{0.1cm}
		log(L) & -1.17$^{+0.06}_{-0.08}$~L${\odot}$\\
		Sloan m$_{g\prime}$ & 16.7$\pm$0.1 \\
		Sloan m$_{r\prime}$ & 15.61$\pm$0.05 \\
		Sloan m$_{i\prime}$ & 14.45$\pm$0.04 \\
		Sloan m$_{z\prime}$ & 14.05$\pm$0.03 \\
		Sloan g-r & 1.1$\pm$0.1 \\
		Sloan r-i & 1.16$\pm$0.07 \\
		\hline
		\hline
		\multicolumn{2}{c}{Astrometry}\\
		\hline
		\multicolumn{2}{c}{2015-05-31}\\
		\hline
		Separation & 240 $\pm$ \changed{42} mas \\
		Position Angle & 85 $\pm$ \changed{13} deg \\
		\hline
		\multicolumn{2}{c}{2022-04-18}\\
		\hline
		Separation & 100.9 $\pm$ 0.7 mas \\
		Position Angle & 145.1 $\pm$ 0.8 deg \\
	    \hline
	\end{tabular}
\end{table}

\section{Observations and Analysis}\label{sec:observationsandanalysis}

\subsection{Observations}
We observed \HIPA\ on April 18th, 2022 with the extreme adaptive optics instrument MagAO-X \citep{Males2022MagAOXSPIE} on the 6.5m Magellan Clay Telescope at Las Campanas Observatory.  We observed \HIPA\ in four science filters: g$^\prime$ ($\lambda_0 = 0.527 \mu$m, $\Delta \lambda_{\rm{eff}} = 0.044 \mu$m), r$^\prime$ ($\lambda_0 = 0.614 \mu$m, $\Delta \lambda_{\rm{eff}} = 0.109 \mu$m), i$^\prime$ ($\lambda_0 = 0.762 \mu$m, $\Delta \lambda_{\rm{eff}} = 0.126 \mu$m), and z$^\prime$ ($\lambda_0 = 0.908 \mu$m, $\Delta \lambda_{\rm{eff}} = 0.130 \mu$m)\footnote{Filter specifications and filter curves can be found in the MagAO-X instrument handbook at \url{https://magao-x.org/docs/handbook/index.html}}.  MagAO-X is equipped with two science cameras, so we carried out science observations in two filters simultaneously.  The science camera EMCCDs were set to 5 MHz readout speed with EM gain 100.  Observations in r$^\prime$, i$^\prime$, and z$^\prime$ had exposure time 0.115~sec; g$^\prime$ had exposure time of 3~sec.  We obtained dark frames of the same settings.  The pixel scale is 6 mas pixel$^{-1}$ (Long et al. in prep), and the science and dark frames were 512$\times$512 pixels (3\arcsec$\times$3\arcsec).  Seeing was stable at 0.4\arcsec\ throughout the observations.

We were unable to obtain observations of a photometric standard star.  We observed HIP~67121 as a photometric standard, only to discover that it is itself a binary with separation too close to resolve but large enough to distort the shape of the PSF core.  We performed all further analysis using \HIPA\ as a photometric reference.

To reduce the raw images in each filter, we dark subtracted each science frame, registered each frame using {\sc photutils DAOStarfinder} \citep{photutils_citation, Stetson1987DAOPHOT} to find the peak of \HIPA\ (uncertainty $\pm$0.05 pixels on peak finding) and {\sc scipy ndimage} \citep{2020SciPy-NMeth} to center it, and rotated each frame to North up and East left (rotate CCW by telescope parallactic angle + 1.995~$\pm$~0.61~deg, Long et al. in prep). Finally we summed the images in each filter to maximize the signal to noise ratio of the faint companion.

Figure \ref{fig:four-filter-image} displays the final images in each science filter, shown with a log stretch.  The companion, \HIPC, is clearly visible at 0.1\arcsec\ to the south east, indicated by the white cross-hairs.  The spacial scale and stretch are the same in each image.  The companion signal was strongest in the z$^\prime$ filter.

\begin{figure*}
\centering
\includegraphics[width=0.95\textwidth]{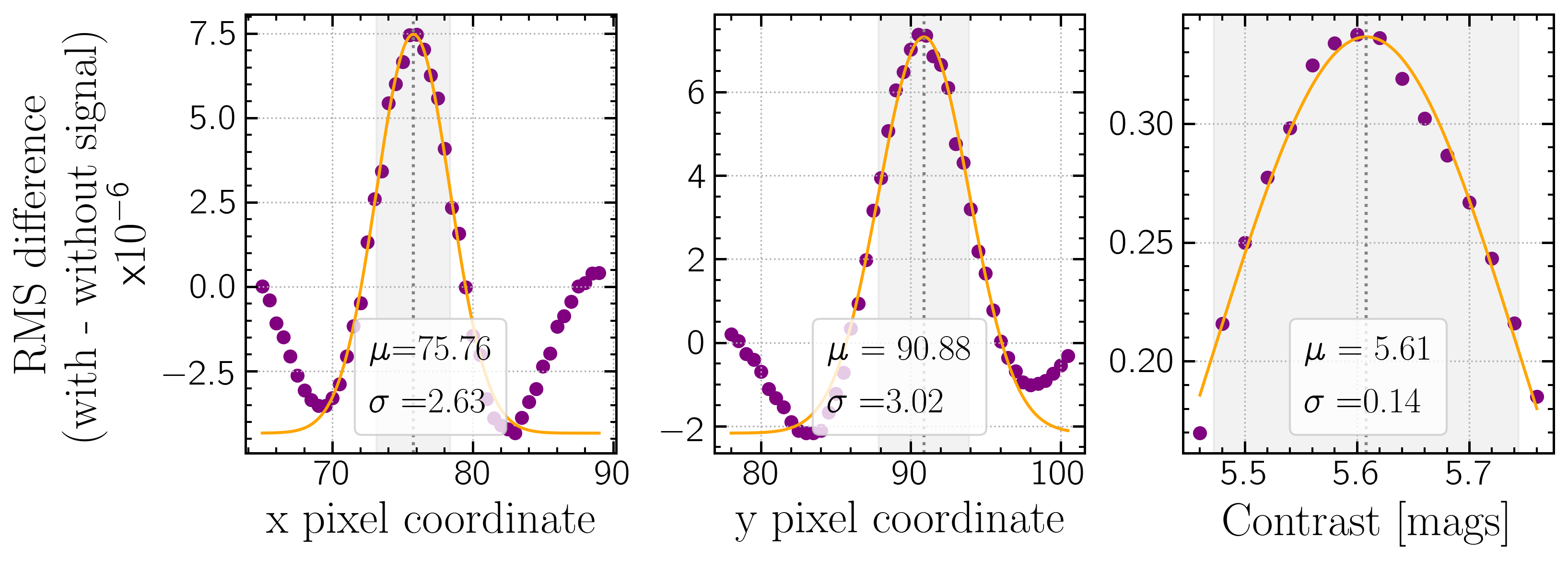}
\caption{Results of our grid search of [$x,y,c$] values for a model which minimizes \HIPC\ residuals post-KLIP processing for the 2015 MagAO/Clio epoch. Each parameter is plotted versus the difference in RMS between KLIP-reduced image with and without the model subtracted. Each parameter was fit with a Gaussian function while keeping the others fixed at their peak value.}
\label{fig:2015-grid-search}
\end{figure*}

\begin{figure}
\centering
\includegraphics[width=0.48\textwidth]{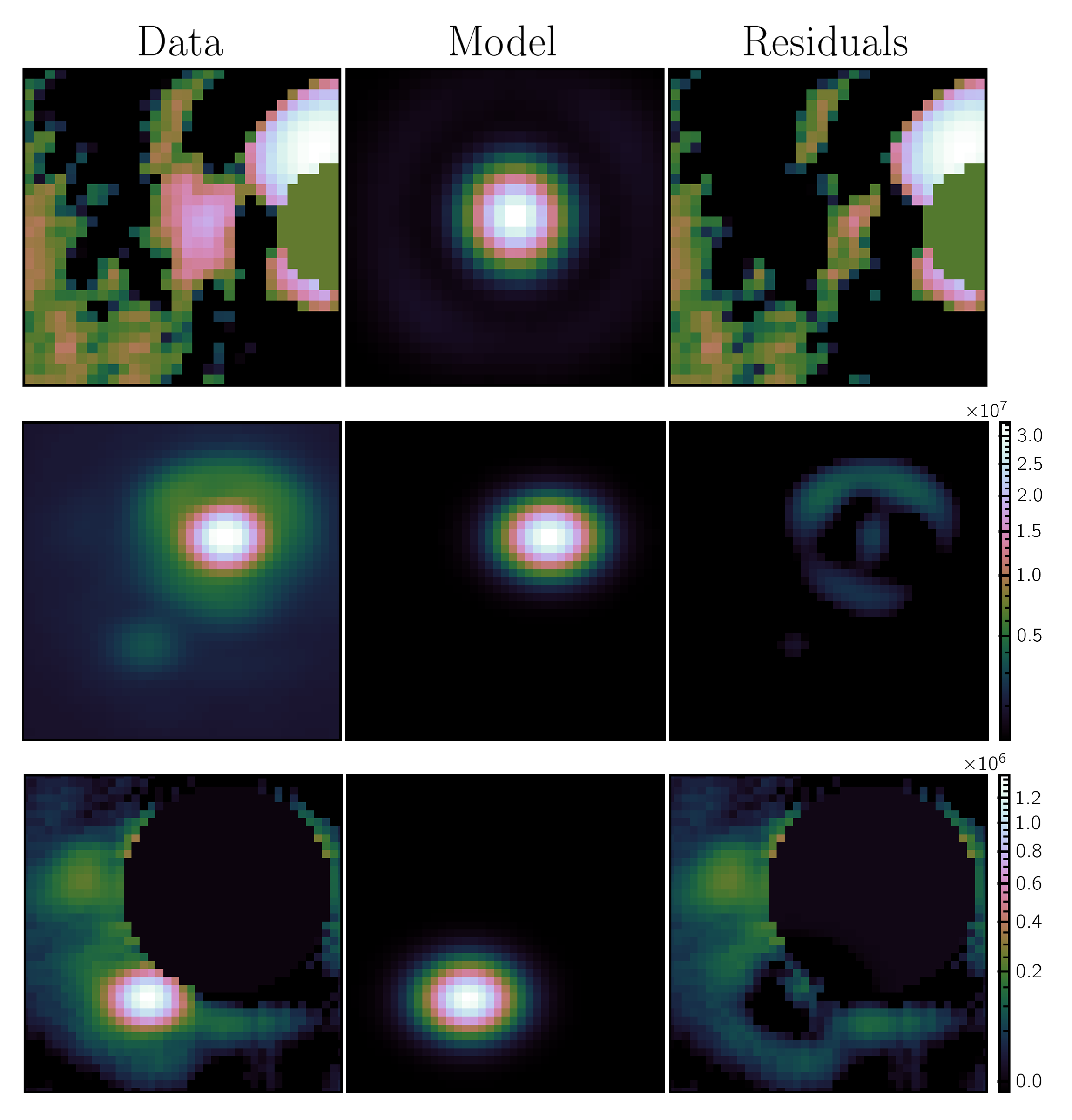}\\
\caption{Top: Data, model, and residual of the [$x,y,c$] that minimizes residuals in 2015 MagAO/Clio observation. Data and residual images are post-KLIP processing, and shown with a log stretch; model image shows the signal with peak values in Figure \ref{fig:2015-grid-search} that was subtracted from images prior to KLIP processing.
Middle and bottom: Data, model, and residuals from the 2D Gaussian model in the 2022 MagAO-X z$^\prime$ image for \HIPA\ (middle) and \HIPC\ (bottom).}
\label{fig:datamodelresiduals}
\end{figure}

\subsection{MagAO-X Photometry}

\textit{Measuring photometry.} We obtained relative photometry for each filter with the following procedure.  We estimated the background level by computing 
the median value in a wide annulus far from the star's halo  (0.6\arcsec-1.2\arcsec).  We used {\sc photutils} aperture photometry tools to sum all pixels in an aperture of radius 1\lod centered on A, and subtracted the sum of pixels with the same aperture area valued at the background level, to estimate the flux from \HIPA.  To estimate the flux from \HIPC\ we repeated the previous with an aperture of the same size centered at its location. We subtracted the mean background value from the image, computed a radial profile of the background subtracted image (excluding the region containing C), and used the flux at C's location in the radial profile to estimate the contribution from \HIPA's halo at that location, and subtracted that as well.  We converted the flux estimates into magnitudes and subtracted to obtain the contrast in MagAO-X filters.

\textit{Uncertainty.} To estimate the uncertainty in the photometry measurements, we used the method of \citealt{mawet_fundamental_2014} for estimating signal to noise ratio in the regime of small number of photometric apertures, as we have at the separation of \HIPC.  At the separation \HIPC, there are N = 2$\pi r$ resolution elements of size \lod (the characteristic scale of speckle noise), where $r = n \lambda$/D and n varies with the filter wavelength. We defined a ring of N-3 resolution elements (neglecting those at and immediately to each side of \HIPC) at separation $r$ with radius 0.5 \lod, then applied Eqn (9) of \cite{mawet_fundamental_2014}, which is the Student's two-sample t-test:
\begin{equation}\label{eq:snr}
    p(x,n2) = \frac{\bar x_1 - \bar x_2}{s_2 \sqrt{1 + \frac{1}{n_2}}}
\end{equation}
where $\bar x_1 = $ \HIPC\ flux, $\bar x_2 =$ mean[$\Sigma$(pixels in apertures)], $s_2 = $ stdev[$\Sigma$(pixels in apertures)], n$_2$ = N-3, and S/N = p.  The denominator of that equation is the noise term.  
We repeated this procedure for \HIPA, defining a ring of apertures beyond the halo of both stars to estimate the background noise.

\textit{Applying the standard.} We used \HIPA\ as the photometric standard star, however literature photometry for \HIPA\ consisted of a blend of flux from \HIPA\ and \HIPC, since it was previously unresolved.  So to use \HIPA\ as a standard we used our measured contrasts to separate the flux contributions from both stars.  First we computed color transformations for MagAO-X filters to Sloan prime system filters using MagAO-X filter curves, public Sloan Digital Sky Survey transmission curves\footnote{\url{http://classic.sdss.org/dr3/instruments/imager/\#filters}}, and a spectral type G5V model from the Pickles Atlas \citep{Pickles1998PicklesModelAtlas}\footnote{MagAO-X to SDSS color transformations for all spectral types can be found in the MagAO-X instrument handbook}.  We obtained published photometry for \HIPA, displayed in Table \ref{tab:HIPA-properties-old-new}, from the UCAC4 catalog \citep{UCAC4-Vizier} and converted to MagAO-X filters using our color transformation.  We then computed the magnitude of \HIPA\ and \HIPC\ in the MagAO-X system as:
\begin{equation}
    A_{\rm{Flux}} + C_{\rm{Flux}} = F_{0,\rm{Vega}}\times 10^{-0.4\times\rm{Total\,mag\,in\,MagAO-X\,system}}
\end{equation}
\begin{equation}
    C_{\rm{Flux}} = A_{\rm{Flux}} \times \rm{Flux\, Contrast}
\end{equation}
\begin{equation}
    A_{\rm{Flux}} \times (1 + 10^{-0.4\times \rm{mag\,Contrast}}) = F_{0,\rm{Vega}}\times 10^{-0.4\times\rm{Total\,mag}}
\end{equation}

We then converted flux of A and C into the Sloan system using color transformation, displaying in Tables \ref{tab:HIPA-properties-old-new} and \ref{tab:HIPC-properties}.

\begin{figure}
\centering
\includegraphics[width=0.48\textwidth]{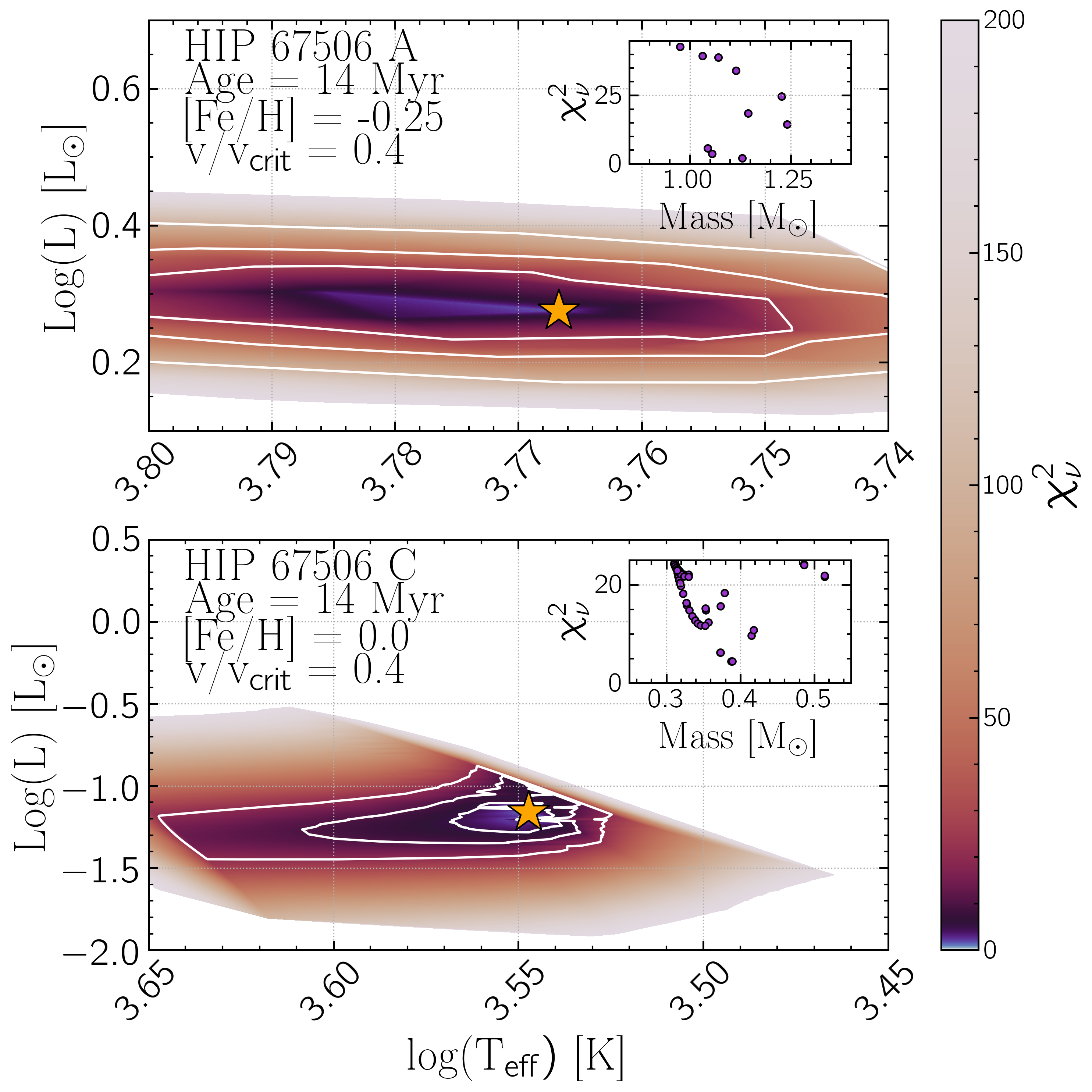}\\
\caption{The lowest $\chi^2$ of all MIST model fits occurred for age$\sim$14 Myr when ages were constrained to be the same for both objects.  This figure shows the map of the reduced $\chi^2$ surface in log(T$_{\rm{eff}}$) and log(L) for \HIPA\ (top) and \HIPC\ (bottom) for age~=~14~Myr and the best-fitting values of metallicity and rotation for each. The lowest reduced $\chi^2$ value for each is marked with an orange star. Contours denote $\chi^2$ = 25, 50, and 100.  Inset axis: $\chi^2$ of model verses model star mass for fits of models with age~=~14~Myr.  The lowest $\chi^2$ values occurred at M${_{\rm{A}}} = 1.13$~\Msun and M${_{\rm{C}}} = 0.39$~\Msun.}
\label{fig:MIST-chi2-map}
\end{figure}

\begin{figure}
\centering
\includegraphics[width=0.49\textwidth]{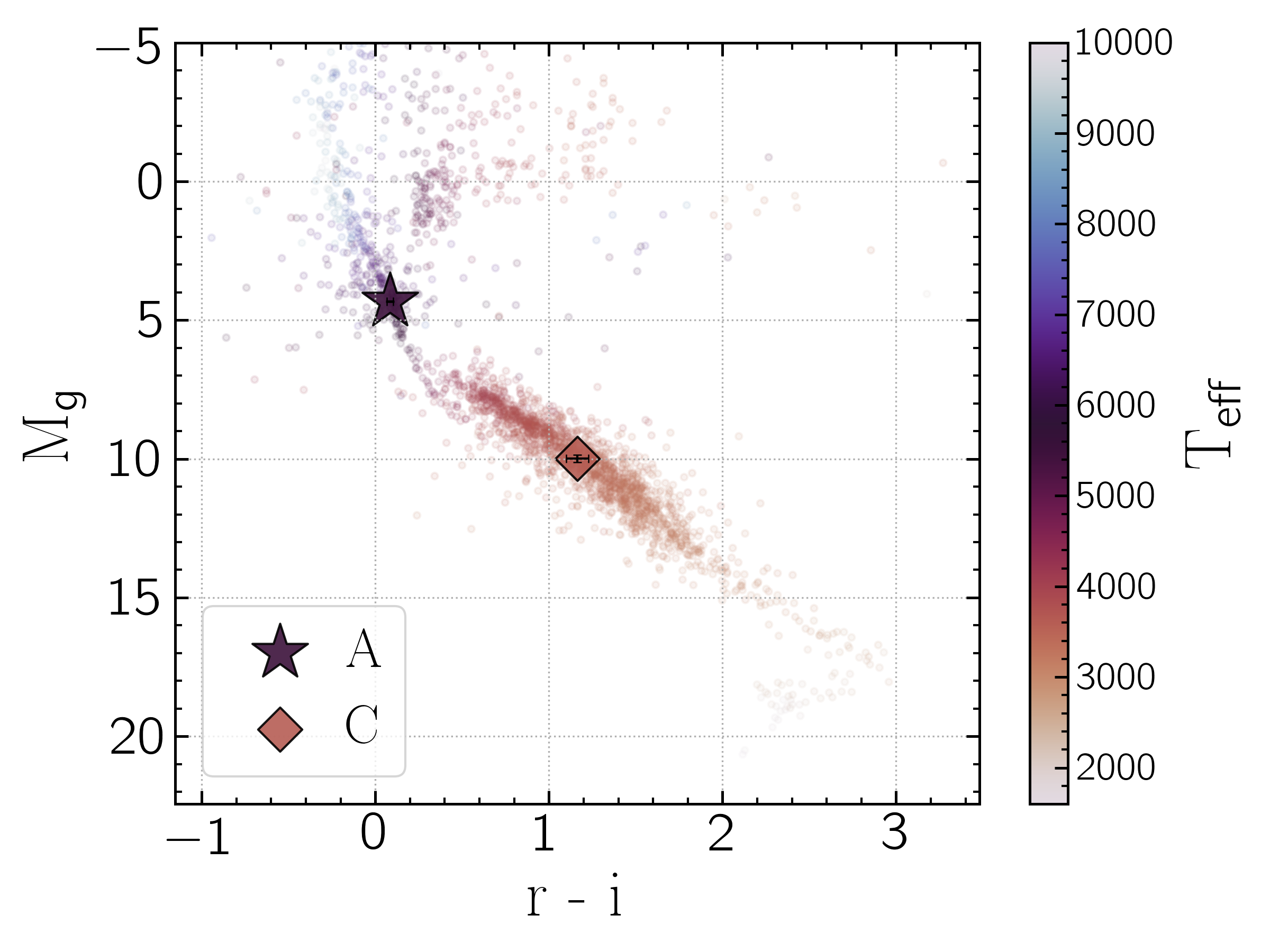}\\
\caption{Color-magnitude diagram (CMD) of Sloan r$^\prime$-i$^\prime$ vs. Sloan g$^\prime$ absolute magnitude.  Points are photometry from the CARMENES sample of well-characterized M- and L dwarfs \citep{Cifuentes2020Carmenes} and a selection of Hipparcos stars with SDSS photometry and T$_{\rm{eff}}$ estimates from \citealt{mcdonald_fundamental_2012}.  Our photometry of \HIPA\ (star) and \HIPC\ (diamond) and uncertainties (black errorbars) are overplotted.  A and C are colored according to the T$_{\rm{eff}}$ of the best-fit MIST model shown in Figure \ref{fig:MIST-chi2-map}.  The best-fitting MIST models correspond to T$_{\rm{eff}}$ values consistent with nearby objects on the CMD.
}
\label{fig:cmd_plot}
\end{figure}

\subsection{Astrometry}

\subsubsection{Relative Astrometry Measurements}
The 2015 MagAO/Clio L$^\prime$ epoch and 2022 MagAO-X epoch give relative astrometry spanning a 7 year baseline.

\textit{The 2015 epoch.} The companion signal has been corrupted by the BDI KLIP algorithm -- it is no longer a recognizable PSF shape, and in \citealt{Pearce2022BDI} we estimated a smaller flux than we measure in this work. The companion signal has been subject to over-subtraction by KLIP, and is not reliable for estimating photometry and astrometry \citep{soummer_detection_2012, Pueyo2016KLIPFM}. 

To estimate the position of the companion, we performed a grid search of the parameters which influence the signal strength in post-processing, similar to \citet{morzinski_magellan_2015} Appendix E.  For a grid of [$x, y$] pixel position and contrast $c$, we injected a negative signal, modeled from the PSF of a median image of the \HIPB\ 2015 dataset, into each \HIPA\ image.  We then performed KLIP reduction via the method in \citealt{Pearce2022BDI} and measured the root-mean-square (RMS) of pixels in a circle of radius 1.5\lod\ ($\sim$11 pixels) centered at the location of the companion signal.

Figure \ref{fig:2015-grid-search} displays the grid search results for the x-pixel coordinate (left), y-pixel coordinate (middle), and contrast (right) versus the difference in RMS between the reduced image with and without the injected signal. We fit a Gaussian to each parameter, while keeping the other parameters fixed at their best value, and took the \changed{mean and standard deviation} as the best modeled parameter.  

Figure \ref{fig:datamodelresiduals} (top) shows the unsubtracted, KLIP-reduced image of \HIPC\ (left, same as Figure \ref{fig:bdi-hipa}, log stretch), the best value model from Figure \ref{fig:2015-grid-search} (middle, linear stretch), and the residuals post-KLIP with that model subtracted from each image pre-KLIP (right, log stretch). With \HIPA\ registered at [$x,y$] = [89.5,89.5] (origin is lower left), we find: $\bar x = 75.76\pm2.63$ pixels, and relative separation \changed{ $\rho_x = 218\pm42$ mas; $\bar y = 90.88\pm3.02$ pixels, $\rho_y = -22\pm48$ mas; total separation and position angle is $\rho = 240\pm42$ mas, $\theta = 85\pm13$ deg.}

\textit{The 2022 epoch.}. We measured the relative astrometry in the MagAO-X z$^{\prime}$ image following a modified version of the method described in \cite{pearce_orbital_2019} and \cite{pearce_boyajians_2021}.  We modeled the PSF core as a simple 2-dimensional Gaussian function and varied the model parameters using the python Markov Chain Monte Carlo package {\sc emcee} \citep{Foreman-Mackey2013emcee} with 100 walkers.  Our model had seven parameters: $x,y$ subpixel position (Gaussian prior with $\mu = $ center from \texttt{DAOStarFinder}, $\sigma = \rm{FWHM}/2.35$, FWHM $=1\lambda$/D at z$^\prime = 0.03$\arcsec), amplitude (Gaussian prior with $\mu = $ peak from \texttt{DAOStarFinder}, $\sigma = $ Poisson noise), background level (Gaussian prior with $\mu = $ mean background level, $\sigma = $ background noise), Gaussian width in the $x$ and $y$ direction (Gaussian prior with $\mu = \rm{FWHM}/2.35$, $\sigma = 0.01$), and rotation relative to x axis (Uniform prior on [0, $\pi$/2]).  The chains converged quickly and we found that 5000 steps was sufficient for chains to converge (Gelman-Rubin statistic $<$ 1.2 for all parameters), with a burn-in of 1000 steps.

We computed the model fit for the location of \HIPA\ and \HIPC\ in the 2022 z$^\prime$ image, where \HIPC's signal was strongest. The data, model, and residuals for the two measurements are shown in Figure \ref{fig:datamodelresiduals} (middle and bottom). We used the MagAO-X astrometric solution (Long et al., in prep)\footnote{Available in the MagAO-X instrument handbook, \url{https://magao-x.org/docs/handbook/}} to compute [$\rho$ (mas), $\theta$ (deg)] for each [$\Delta x$,$\Delta y$] (pixels) between A and C in the MCMC chains, then took the mean and standard deviation as the [$\rho, \theta$] for the 2022 epoch. Detector distortion is negligible at 0.1\arcsec\ (Long et al. in prep).
We find $\rho = 100.9 \pm 0.7$~mas, $\theta = 145.1 \pm 0.8 $~deg.

\section{Results}\label{sec:results}

\subsection{Photometry}\label{sec:photometry}
We compared our magnitudes in the Sloan filter system with synthetic photometry from two stellar evolution grids, the Mesa Isochrones and Stellar Tracks \citep[MIST, ][]{Dotter2016MIST,Choi2016MIST,Paxton2011MESA,Paxton2013MESA,Paxton2015MESA},  
and stellar tracks and isochrones with the Padova and Trieste Stellar Evolution Code \citep[PARSEC, ][]{Bressan2012}.

We used our absolute g$^\prime$, r$^\prime$, i$^\prime$, and z$^\prime$ SDSS magnitudes for \HIPA\ and \HIPC\, as well as g$^\prime$-r$^\prime$ and r$^\prime$-i$^\prime$ colors for evaluating which models in each grid best describe our observations.  For each isochrone set we minimized the $\chi^2$ of the synthetic photometry to our data as
\begin{equation}
    \chi^2 = \sum { \left( \frac{M_{\rm{x,obs}} - M_{\rm{x,model}}}{M_{\rm{x,uncert}}} \right)^2 }
\end{equation}
where $M_{\rm{x}}$ is the absolute magnitude in a given filter or $\Delta$ magnitude in a color.  We imposed the constraint that the age must be the same for \HIPA\ and \HIPC, and computed the final goodness of fit as $\chi^2 = \chi^2_{A} + \chi^2_{C}$.  

We obtained the MIST\footnote{Accessed from \url{https://waps.cfa.harvard.edu/MIST/model\_grids.html}} isochrone synthetic photometry in the SDSS ugriz system with rotation rate $v/v_{\rm{crit}} = 0.0$ and 0.4, [Fe/H] = [-4.00, -2.00] in 0.50 dex steps and [Fe/H] = [-2.00, +0.50] in 0.25 dex steps, and log(Age) = [5.0, 10.3] in 0.05 dex steps.

For MIST isochrone $\chi^2$ minimization, we determine T$_{\rm{eff}}$~=~6000$\pm$350~K and log(L)~=~0.28$^{+0.06}_{-0.08}$~L$_\odot$ for \HIPA, T$_{\rm{eff}}$~=~3600$^{+250}_{-350}~K$ and log(L)~=~-1.17$^{+0.06}_{-0.08}$~L$_\odot$ for \HIPC.  

Figure \ref{fig:MIST-chi2-map} shows the reduced $\chi^2$ surface for log(T$_{\rm{eff}}$) and log(L) for the overall lowest $\chi^2$ MIST isochrone ($\chi^2 = 36.7$), with age =~14~Myr, rotation v/v$_{\rm{crit}}$~=~0.4, and [Fe/H]~=~0.25 for A and [Fe/H]~=~0.0 for C.  Values of log(T$_{\rm{eff}}$) are not well constrained for A, spanning from log(T$_{\rm{eff}}$)$\sim$3.76--3.78 (5700--6000K).  The insets in Figure \ref{fig:MIST-chi2-map} display reduced $\chi^2$ as a function of mass at 14~Myr, with the best fitting values occurring at M$_{\rm{A}}$~=~1.1\Msun, M$_{\rm{C}}$~=~0.4\Msun.  A second local minimum ($\chi^2 = 39.2$) occurred at age~=~5.6~Gyr, M$_{\rm{A}}$~=~1.1\Msun, and M$_{\rm{C}}$~=~0.65\Msun.  (\textcolor{black}{A plot of $\chi_{\rm{min}}^2$ as a function of age is included in the supplementary material.})

\begin{figure}
\centering
\includegraphics[width=0.49\textwidth]{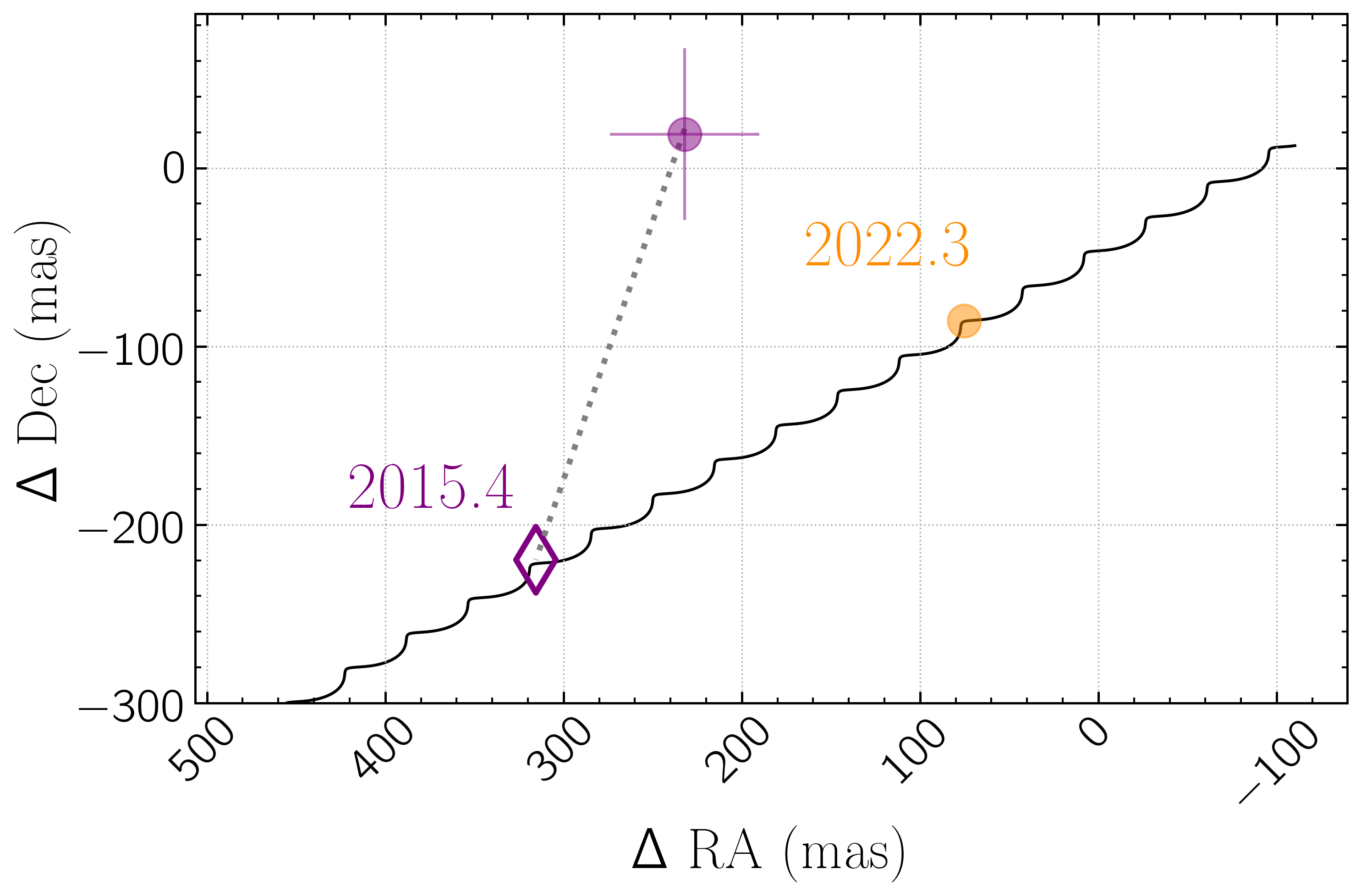}\\
\caption{Relative astrometry of \HIPC\ relative to A for the MagAO 2015 epoch (purple) and the MagAO-X 2022 epoch (orange).  The abscissa and ordinate axes display position of \HIPC\ relative to A in mas in right ascension (RA) and declination (Dec).  The motion of a non-moving background object at the position of \HIPC\ is given by the black track; the predicted position in 2022, given then 2015 position, is an open diamond.  The observed position and uncertainty in each epoch is shown as filled circles (uncertainties are smaller than the marker for the 2022 epoch).  The observed motion of the \HIPC\ is not consistent with a background object, and is likely due to orbital motion.
}
\label{fig:cpm_plot}
\end{figure}

We used PARSEC version 1.2S\footnote{Accessed from \url{http://stev.oapd.inaf.it/cgi-bin/cmd}} with the YBC bolometric correction library \citep{Chen2019YBC} and revised Vega SED from \citet{Bohlin2020UpdatedHSTFluxCal}, and retrieved isochrone tables from log(age) = [6.0, 10.13] dex in intervals of 0.1 dex and metalicities [M/H] = [-4.0, 0.5] dex in intervals of 0.5 dex, with synthetic photometry in the SDSS ugriz system. For PARSEC isochrone $\chi^2$ minimization, we determine T$_{\rm{eff}}$~=~6000$\pm$350~K and log(L)~=~0.29$^{+0.06}_{-0.08}$~L$_\odot$ for \HIPA, T$_{\rm{eff}}$~=~3600$^{+250}_{-350}~K$ and log(L)~=~-1.18$^{+0.06}_{-0.08}$~L$_\odot$ for \HIPC.    \changed{Our photometry was insufficient to place meaningful constraints on the age of either star. }


Figure \ref{fig:cmd_plot} shows a color-magnitude diagram of SDSS r-i color vs. SDSS g absolute magnitude.  \HIPA\ (purple star) and \HIPC\ (orange diamond) are plotted with our photometry and colored according to our isochrone-derived T$_{\rm{eff}}$ estimates.  Also plotted are reference stars from the CARMENES sample of well-characterized M- and L dwarfs \citep{Cifuentes2020Carmenes} and a selection of Hipparcos stars with SDSS photometry and T$_{\rm{eff}}$ estimates from \citealt{mcdonald_fundamental_2012}.  Our colors and temperature estimates are consistent with the reference stars. 
We estimate the spectral type of \HIPA\ and \HIPC\ to be SpT$_{\rm{A}}$ $\approx$ F8V--G2V and  SpT$_{\rm{C}}$ $\approx$ K7V--M2V.

\subsection{Astrometry}\label{sec:astrometry}

Figure \ref{fig:cpm_plot} displays a common proper motion plot of \HIPC\ relative to \HIPA. We show the observed separation of \HIPC\ in right ascension and declination for the 2015 and 2022 epochs (filled circles and error bars), the expected track if \HIPC\ were a non-moving background object (zero proper motion; black track), and the predicted position of \HIPC\ at the 2015 observation if it were a background object (open diamond).  The observed position of \HIPC\ does not follow the expected motion for a distant background object.  We infer that the relative motion of \HIPC\ is more consistent with a bound object than an unassociated object.  This is supported by the large proper motion anomaly of \HIPA.

\changed{Using the two position angles of Table \ref{tab:HIPC-properties}, we determined that the position angle of \HIPC\ at the Gaia epoch of 2016.0 was 90$\pm$12$^{\circ}$, which agrees with the proper motion anomaly vector PA at the Gaia epoch of 96.6$\pm$4.1$^{\circ}$ \citep{Kervella2022}.
}

\changed{Our astrometry was insufficient to meaningfully constrain the orbit or dynamical mass, due to there being only two astrometric points and large error bars on the 2015 epoch.
}

\section{Conclusion}\label{sec:conclusion}
\changed{We have shown that \HIPA\ has a previously unknown 0.1\arcsec\ companion, originally detected in 2015 with MagAO/Clio and BDI in L$^{\prime}$. The shape was distorted from a typical PSF due to post-processing, and might have been easily dismissed with the other residuals at that radius. However several secondary indications hinted that the dubious candidate companion signal for \HIPA\ in \citet{Pearce2022BDI} was a strong candidate and merited follow-up observations: the poor Gaia astrometric signal, the significant PMa \changed{with the right acceleration vector angle}, and the overluminosity of the Gaia photometry. Our analysis in \citealt{Pearce2022BDI} pointed to a possible high mass brown dwarf. We followed up in 2022 with MagAO-X and the companion was immediately and easily detected and determined to be a low mass star. The low S/N signal of \HIPC\ at such a small IWA was bolstered by secondary indicators, which turned out to be powerful predictors of the genuine companion.  We estimate \HIPA\ and \HIPC\ to be type F8--G2 and K7--M2 respectively.  Further astrometric and photometric measurements are required to constrain properties and orbital elements. 
}

\section*{Acknowledgements}

L.A.P.~acknowledges research support from the NSF Graduate Research Fellowship.  This material is based upon work supported by the National Science Foundation Graduate Research Fellowship Program under Grant No. DGE-1746060. 

J.D.L.~thanks the Heising-Simons Foundation (Grant \#2020-1824) and NSF AST (\#1625441, MagAO-X).

S.Y.H~was supported by NASA through the NASA Hubble Fellowship grant \#HST-HF2-51436.001-A awarded by the Space Telescope Science Institute, which is operated by the Association of Universities for Research in Astronomy, Incorporated, under NASA contract NAS5-26555.

MagAO-X was developed with support from the NSF MRI Award \#1625441. The Phase II upgrade program is made possible by the generous support of the Heising-Simons Foundation.

We thank the LCO and Magellan staffs for their outstanding assistance throughout our commissioning runs.

This work has made use of data from the European Space Agency (ESA) mission
{\it Gaia} (\url{https://www.cosmos.esa.int/gaia}), processed by the {\it Gaia}
Data Processing and Analysis Consortium (DPAC,
\url{https://www.cosmos.esa.int/web/gaia/dpac/consortium}). Funding for the DPAC
has been provided by national institutions, in particular the institutions
participating in the {\it Gaia} Multilateral Agreement.

This research has made use of the Washington Double Star Catalog maintained at the U.S. Naval Observatory.

\vspace{5mm}
\textit{Facilities:} Las Campanas Observatory, Magellan:Clay (MagAO-X)

\textit{Software:}Numpy \citep{Harris2020Numpy}, Astropy \citep{astropy:2018}, Matplotlib \citep{Hunter:2007}, Scipy \citep{2020SciPy-NMeth}, emcee \citep{Foreman-Mackey2013emcee}, corner.py \citep{foreman-mackey_cornerpy_2016}, Photutils \citep{photutils_citation}

\section*{Data Availability}


The data underlying this article are available in at \url{https://github.com/logan-pearce/HIP67506-AC-Public-Data-Release} and at DOI: 10.5281/zenodo.7098006.



\bibliographystyle{mnras}
\bibliography{references} 




\appendix

\appendix
\section{\HIPB\ is not a wide binary companion to \HIPA}\label{appendixA}

The Gaia solutions for \HIPA\ and \HIPB\ show differing parallax solutions (A: source id~=~6109011780753115776, $\pi$~=~4.51 mas; B: source id~=~6109011742094383744, $\pi$~=~0.55 mas), indicating that \HIPB\ is an order of magnitude more distant than \HIPA. 
This raises the question if the two stars are actually a gravitationally bound pair versus a chance alignment of unassociated stars at different distances. 
We queried the Gaia catalog for all objects within a 1$^{\circ}$ radius of \HIPA\ and used a simple Monte Carlo simulation to determine that, given the density of objects in the local region, the probability of a chance alignment of two stars within a 9\arcsec\ radius is 38.9~$\pm$~1.6\%.  The probability of chance alignment of two stars within 9\arcsec\ and 2 magnitudes is 4.5~$\pm$~0.7\%. So it is plausible that they are a chance alignment.

\begin{figure*}
\centering
\includegraphics[width=0.7\textwidth]{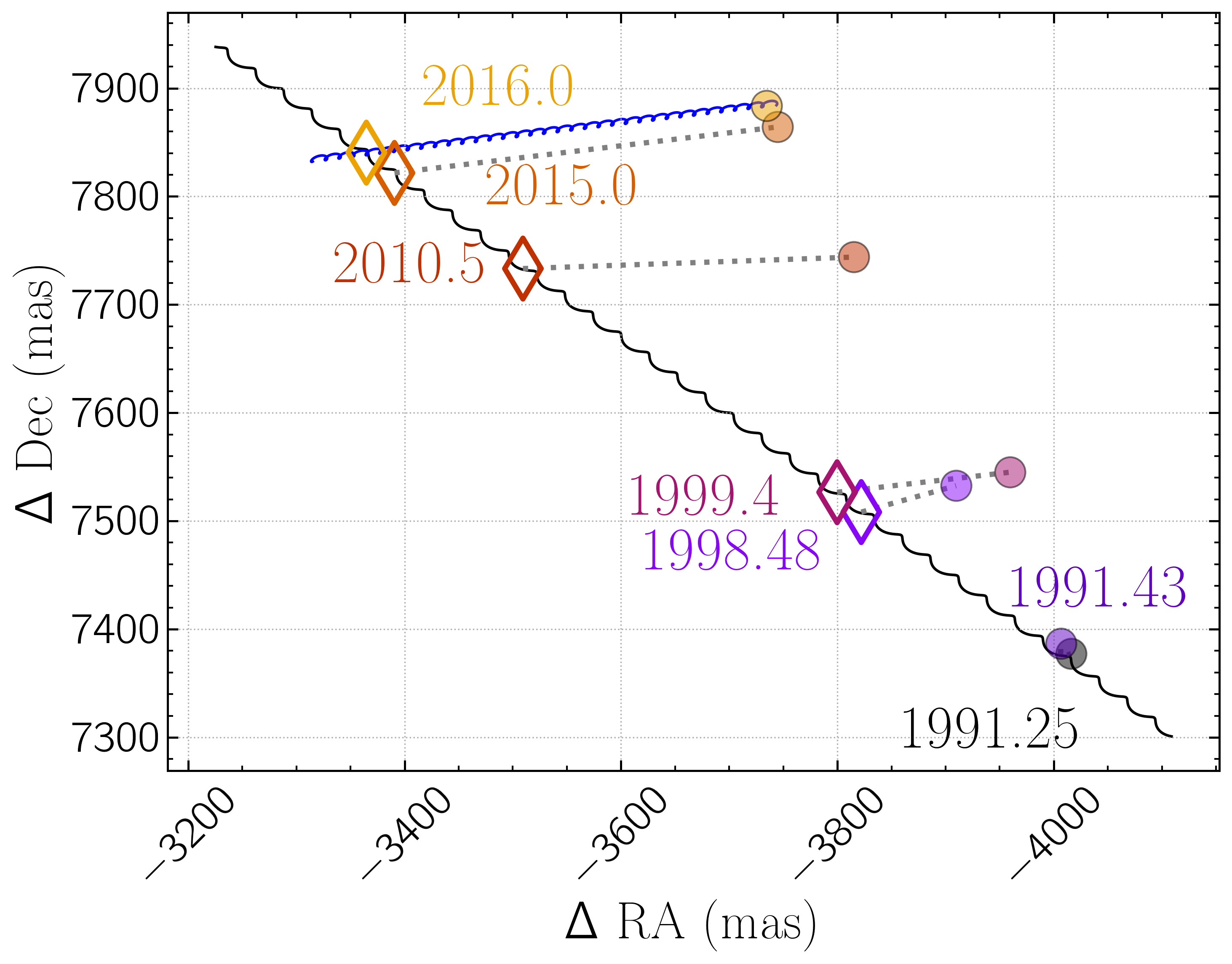}
\caption{Relative astrometry of \HIPA\ and \HIPB\ (WDS~J13500-4303~A and B).  The abscissa and ordinate axes display position of \HIPB\ relative to \HIPA\ in mas in right ascension (RA) and declination (Dec).  The motion of a non-moving background object at the position of \HIPB\ is shown by the black track for the Gaia DR3 proper motion and parallax given for \HIPB, with the predicted position at WDS observation epochs marked by colored diamonds.  The blue track shows the track over the same time span given by the Gaia DR3 proper motion and parallax of \HIPB.  The observed WDS positions shown in Table \ref{tab:TYC-WDS_astr} are marked by filled circles with corresponding epoch colors. 
The observed motion of \HIPB\ relative to \HIPA\ is consistent with the Gaia DR3 proper motion and not with a common proper motion pair. We conclude that the order-of-magnitude higher distance for \HIPB\ than \HIPA\ given by Gaia DR3 is correct.
}
\label{fig:BDI1350CPM}
\end{figure*}

The Washington Double Star Catalog \citep[WDS; ][]{Mason2001WDS} astrometry for this system (WDS J13500-4303 A and B) is shown in Table \ref{tab:TYC-WDS_astr}.  Figure \ref{fig:BDI1350CPM} displays the motion of \HIPB\ relative to \HIPA\ as observed in WDS (circles), the predicted position of \HIPB\ if it were an unmoving background star and \HIPA\ moved with the proper motion given by Gaia DR3 (black track and diamonds), and the Gaia DR3 proper motion and parallax track for \HIPB\ (blue track).  The WDS astrometry is consistent with the Gaia proper motion and parallax and not a gravitationally bound pair with common proper motion, indicating that the small parallax in Gaia DR3 for \HIPB\ is correct and the two are unassociated.

Assuming a mass of 1.2~M$_\odot$ for both stars (since \HIPB\ appears to have a similar brightness as A), the escape velocity at the current separation is 1.306~$\pm$~0.005~km~s$^{-1}$. Taking the case of a face-on orbit (radial velocity = 0~km~s$^{-1}$, the smallest possible value for the relative velocity vector), the observed linear motion shown in Figure \ref{fig:BDI1350CPM} gives a velocity of 24$\pm$2~km~s$^{-1}$, roughly 14-$\sigma$ larger than the escape velocity. \cite{Clarke2020GaiaBinaryPMs} and \cite{Belokurov2020UnresolvedBinariesDR2} showed that unresolved hierarchical triples and high RUWE astrometric solutions can produce relative velocities exceeding escape velocity and an apparent deviation from Newtonian gravity in the case of bound systems,  
so we are unable to entirely rule out their being a gravitationally bound system.  But the remarkable agreement of WDS astrometry with the Gaia proper motion solutions strongly favors the Gaia parallaxes and proper motions being accurate.

We conclude that the two sources are not a gravitationally bound system, and that the star \HIPB\ is not in fact a companion to \HIPA, but a much further distant background star.  

\begin{table*}
	\centering
	\caption{WDS catalog entry for \HIPA\ and \HIPB\ (WDS~J13500-4303~A~and~B)}
	\label{tab:TYC-WDS_astr}
	\begin{tabular}{cccccc} 
		\hline
		Date & Position Angle & PA Error & Sep & Sep Error &  Ref\\
		 &  (deg) & (deg) & (arcsec) & (arcsec) & \\
		\hline
		1991.25 & 323.3 & - & 9.190 & - & \citealt{Hip1997Vizier} \\
        1991.43 & 323.4 & - & 9.19 & - &  \citealt{Fabricius2002TychoDoubleStarCat} \\
        1998.482 & 324.6 & 0.1 & 9.230 & 0.001 & \citealt{Hartkopf2013UCACDoubleStarCat} \\
        1999.40 & 324.3 & - & 9.28 & - &  \citealt{Cutri2003-2MASSVizier} \\
        2010.5  & 326.0 & 0.9 & 9.33 & 0.15 &  \citealt{Cutri2012WISE} \\
        2015.0  & 326.899 & - & 9.377 & - & \citealt{Knapp2018EstimatingVisMagsForWDSStars} \\ 
        2016.0 & 327.0363	& 0.0002 & 9.38593 & 3e-05 & \citealt{gaiaEDR3} \\
		\hline
	\end{tabular}
\end{table*}



\bsp	
\label{lastpage}
\end{document}